\documentclass{webofc}
\usepackage[varg]{txfonts}   
\begin{document}
\title{IceCube Sterile Neutrino Searches}
%
%

\author{\firstname{B. J. P. } \lastname{Jones}\inst{1}\fnsep\thanks{\email{ben.jones@uta.edu}}
        \firstname{For the IceCube Collaboration}
}

\institute{University of Texas at Arlington, 108 Science Hall, 502 Yates St, Arlington, TX
          }

\abstract{%
  Anomalies in short baseline experiments have been interpreted as evidence for additional neutrino mass states with large mass splittings from the known, active flavors.  This explanation mandates a corresponding signature in the muon neutrino disappearance channel, which has yet to be observed.  Searches for muon neutrino disappearance at the IceCube neutrino telescope presently provide the strongest limits in the space of mixing angles for eV-scale sterile neutrinos.  This proceeding for the Very Large Volume Neutrino Telescopes (VLVnT) Workshop summarizes the IceCube analyses that have searched for sterile neutrinos and describes ongoing work toward enhanced, high-statistics sterile neutrino searches.
}
\maketitle
\section{Introduction}
\label{intro}
Sterile neutrinos are hypothetical particles that have been invoked to explain anomalies in short baseline accelerator decay-at-rest \cite{Athanassopoulos:1996jb} and decay-in-flight experiments \cite{Aguilar-Arevalo:2018gpe}, reactor neutrino fluxes \cite{Serebrov:2018vdw,Mention:2011rk} and radioactive source experiments \cite{Giunti:2010zu}.   In low-energy, short baseline experiments that are sensitive to $\nu_e$ appearance, matter effects can be neglected and an oscillation of the form:
\begin{equation}
    P_{\nu_\mu\rightarrow\nu_e}=\sin^2 2\theta_{\mu e}\sin^2\left[\frac{\Delta m^2 L}{4E}\right]
\end{equation}
is predicted.  A large $\Delta m^2$ thus introduces an oscillation at a small characteristic $L/E$, with the effective mixing parameter, $\sin^2 2\theta_{\mu e}$ governing the amplitude of oscillation. To introduce flavor-change at similar $L/E$ as exhibited in the MiniBooNE and LSND experiments, mixing parameters $\sin^2 2\theta_{\mu e}$ of 10$^{-3}$ or larger are required, with favored parameter space in the one-to-few eV$^2$ mass splittings. \cite{Kopp:2013vaa,Conrad:2012qt,Gariazzo:2017fdh}.

Oscillations in experiments sensitive to disappearance signatures exhibit a similar functional form but with a different effective mixing parameters. Oscillation probabilities in the absence of matter effects take the form:
\begin{equation}
    P_{\nu_\alpha\rightarrow\nu_\alpha}=1-\sin^2 2\theta_{\alpha\alpha}\sin^2\left[\frac{\Delta m^2 L}{4E}\right],
\end{equation}
Where $\alpha$ is the disappearing flavor. To explain apparent anomalies in disappearance experiments, mixing parameters $\sin^2 2\theta_{ee}$ of $\mathcal{O}(0.1)$ and $\Delta m^2$ values of larger than $\sim$0.3 eV$^2$ are required \cite{Giunti:2017yid}.  

In the minimal scenario with a single heavy sterile neutrino (3+1), the effective mixing parameters in vacuum-like oscillation experiments $\sin^2 2\theta_{\mu e}$ and $\sin^2 2\theta_{ee}$ can be related to elements of an extended leptonic mixing matrix via:
\begin{equation}
    \sin^2 2\theta_{ee}=4 |U_{e4}|^2(1-|U_{e4}|^2),\quad
    \sin^2 2\theta_{\mu \mu}=4 |U_{\mu 4}|^2(1-|U_{\mu 4}|^2),\quad
    \sin^2 2\theta_{\mu e}=4 |U_{\mu 4}|^2|U_{e4}|^2
\end{equation}
A finite $\nu_\mu\rightarrow\nu_e$ appearance signature implies a finite value for both $\sin^2 2\theta_{\mu \mu}$ and  $\sin^2 2\theta_{ee}$.  A generic prediction of sterile neutrino models that explain short baseline $\nu_\mu\rightarrow\nu_e$ appearance anomalies is that there should be a finite disappearance signature in the channel $\nu_\mu\rightarrow\nu_\mu$.

\begin{figure}[t]
\begin{center}
\includegraphics[width=0.99\columnwidth]{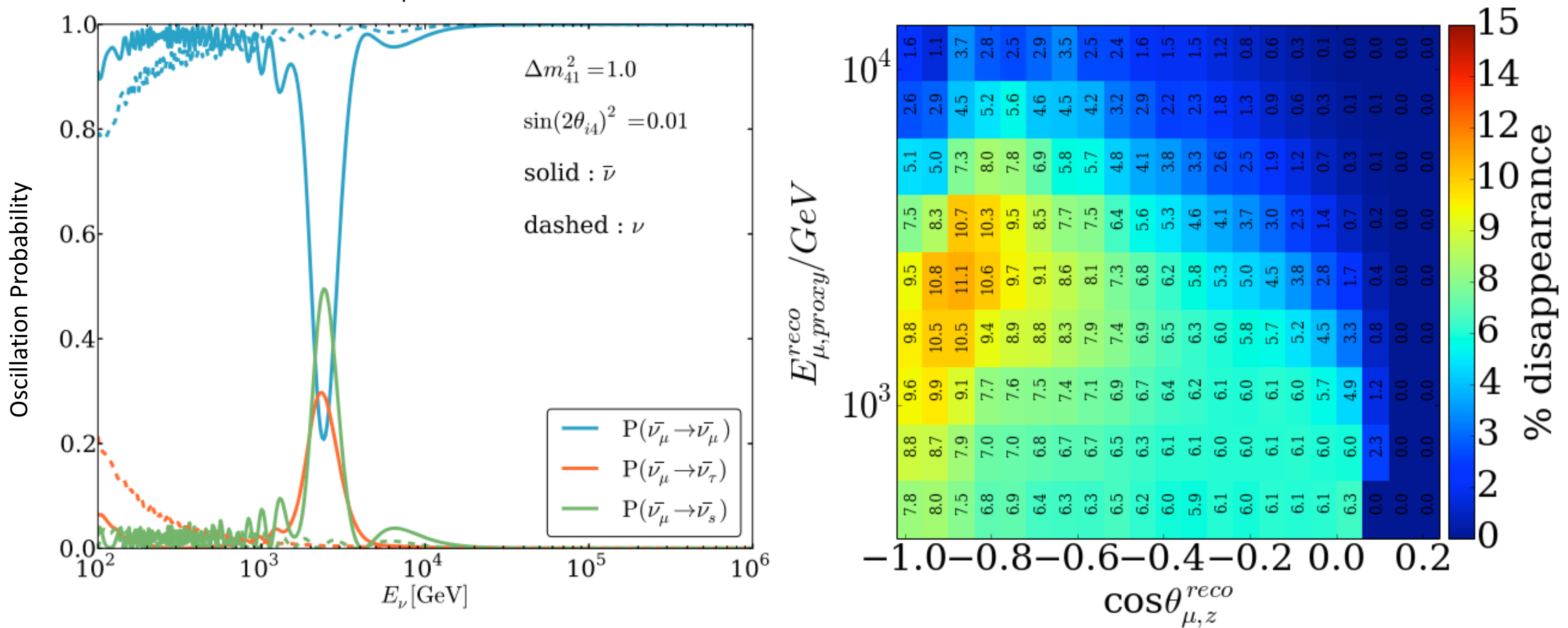}
 
\caption{Left: theoretical oscillation probabilities displaying matter resonance; Right: observable signature of this parameter point in the IceCube high-energy sterile neutrino search.\label{fig:OscProb}}
\end{center}
\end{figure}

Muon neutrino disappearance can be probed by atmospheric neutrino oscillation experiments such as SuperKamiokande \cite{Abe:2014gda} and IceCube \cite{TheIceCube:2016oqi,Aartsen:2017bap}, as well as accelerator neutrino experiments \cite{MINOS:2016viw,Adamson:2017uda}.  Of these, IceCube probes the highest energy range, with a high-statistics sample of well-reconstructed atmospheric neutrinos spanning the range 6 GeV to 20 TeV.  This is a regime where matter effects are not only non-negligible, but can be very large.  For a 1~eV$^2$ sterile neutrino, for example, a large matter-induced resonance would be expected at 3 TeV, greatly amplifying the ordinarily small oscillation probability \cite{Choubey:2007ji,Barger:2011rc,Esmaili:2013cja,Esmaili:2013vza,Lindner:2015iaa}.  An example oscillation spectrum calculated by the {\tt NuSQUIDS} software package \cite{Delgado:2014kpa} is shown in Fig \ref{fig:OscProb}.

IceCube has performed searches for sterile neutrinos in both high-energy \cite{TheIceCube:2016oqi} and low-energy \cite{Aartsen:2017bap} data samples. No evidence for anomalous muon neutrino disappearance was observed in either case.  These null observations, along with disappearance constraints from other experiments, have introduced a severe tension into sterile neutrino models.  The viable parameter space for a simple sterile neutrino explanation of short baseline and $\nu_e$ disappearance anomalies is now small according to some commentators \cite{Giunti:2019aiy} and vanishing according to others \cite{maltoni_michele_2018_1287015}.  This has prompted consideration of more exotic scenarios to explain the anomalies, which may \cite{Liao:2018mbg,Ballett:2018ynz,Bai:2015ztj,Moss:2017pur} or may not \cite{Asaadi:2017bhx,Arguelles:2018mtc,Jordan:2018qiy,Doring:2018cob,Bertuzzo:2018itn} include sterile neutrinos.  A new high-energy sterile neutrino analysis from IceCube using seven years of data and commensurately improved control of systematic uncertainties is now in preparation. This analysis will probe the region of small mixing angles to further constrain the parameter space of sterile neutrino models.  

In this proceeding we briefly describe the one-year IceCube high-energy sterile neutrino analysis (Sec.~\ref{sec-1}), the three-year IceCube low-energy sterile neutrino analysis (Sec.~~\ref{sec-2}), and the forth-coming seven-year IceCube sterile neutrino search (Sec.~\ref{sec-3}).

\section{IceCube high-energy Sterile Neutrino Search}
\label{sec-1}

The high-energy sterile neutrino search at IceCube \cite{TheIceCube:2016oqi}  used one year of atmospheric neutrino data passing an event selection developed for a search for diffuse astrophysical muon neutrinos \cite{Aartsen:2015rwa}. The up-going tracks from atmospheric neutrinos dominate below $\sim$100~TeV, and have a particularly clean signature, with the Earth acting as a filter against contamination from muons created in cosmic ray air showers. The sample is thus effectively background-free.  The dataset used to search for resonant sterile neutrino oscillations contains 20,145 reconstructed up-going muon tracks in the approximate energy range 320~GeV to 20~TeV.

In order to remain sub-dominant to statistical uncertainty, systematic uncertainties in the shape of the reconstructed spectrum must be controlled at the level of around  6-7\% per bin.  The dominant uncertainties include the properties of the South Pole ice~\cite{Aartsen:2013rt}, efficiency of the digital optical modules~\cite{Hanson:2006bk}, and the atmospheric neutrino flux shape, which was parameterized by a tunable spectral index, $\bar\nu/\nu$ ratio, $\pi / K$ production ratio and a set of discrete primary cosmic ray models propagated via the MCEq cascade calculation \cite{Fedynitch:2015zma,Fedynitch:2012fs}.  Additional sources of systematic uncertainty including neutrino cross section, Earth density model and atmospheric density profile were also incorporated, but shown to be sub-dominant.

\begin{figure}[t]
\begin{center}
\includegraphics[width=0.6\columnwidth]{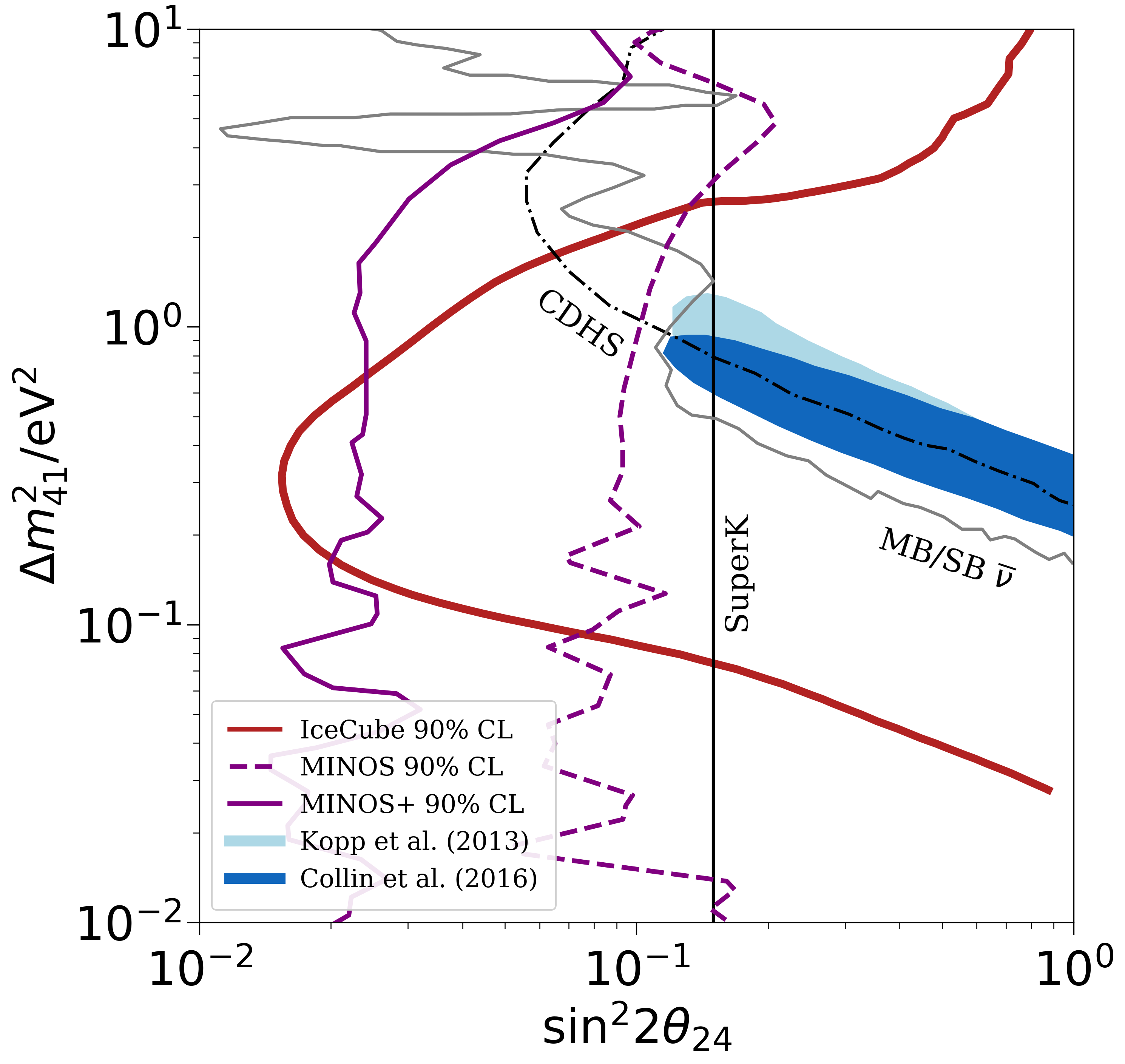}
 
\caption{90\% CL limit from the IceCube high-energy sterile neutrino search compared with allowed regions from appearance experiments (blue) and limits at the time of publication (grey/black), as well as subsequent $\nu_\mu$ disappearance results published since (purple). \label{fig:HEResult}}
\end{center}
\end{figure}

No evidence for oscillation was found within experimental sensitivity.  This places a strong constraint on the mixing angle $\sin ^2 2\theta_{24}$ extending to 0.02 at 0.3 eV$^2$.  This limit on $\theta_{24}$ is constructed with the conservative choice of $\theta_{34}=0$. A stronger limit in $\sin ^2 2\theta_{24}$ is implied for non-zero $\theta_{34}$ \cite{Lindner:2015iaa}.   The negative IceCube result, compared to other negative results obtained from searches for $\nu_\mu$ disappearance experiments as well as and the allowed region from appearance experiments at the time of the IceCube publication are shown in Fig~\ref{fig:HEResult}.

\section{IceCube Low-energy Sterile Neutrino Search}
\label{sec-2}

At low energies (<100 GeV) IceCube is sensitive to standard atmospheric neutrino oscillations \cite{Aartsen:2017nmd}.  The inclusion of sterile neutrino mixing within an extended neutral lepton mixing matrix impacts the oscillation probability in this region \cite{Razzaque:2012tp}, with scale of effect proportional to the matter density traversed.  In IceCube, the observable effect is independent of $\Delta m^2$ within the mass range of interest, since the oscillations are fast enough to be averaged by the energy resolution of the detector.  The most pronounced effect is expected at an energy of 20 GeV for upgoing muons \cite{Aartsen:2017bap}.  

An event selection was developed to isolate muon neutrino events between 6.3 and 56~GeV, and applied to three years of IceCube data to yield 5,118 total events. Because of their low energies, reconstruction is more challenging and background rejection more difficult than in the higher energy sample. To mitigate against backgrounds, the DeepCore sub-array was used for event selection and reconstruction with the remainder of the IceCube array serving as a veto against atmospheric muon backgrounds.

At these energies the properties of the refrozen ice in the immediate vicinity of the detector strings dominates the ice uncertainty budget.  Along with digital optical module efficiency, this represents the largest detector systematic uncertainty.  low-energy neutrino interactions require control of distinct cross section uncertainties to the high-energy, deep inelastic samples, including the resonant and quasielastic axial masses.  Flux parameter uncertainties including spectral index and an energy dependent $\nu/\bar \nu$ flux ratio are included.  Finally, since lower energy muons are more challenging to reconstruct and select than their higher energy counterparts, $\nu_e$ and atmospheric muon contamination in the sample is calculated and parameterized with an uncertainty.

\begin{figure}[t]
\begin{center}
\includegraphics[width=0.65\columnwidth]{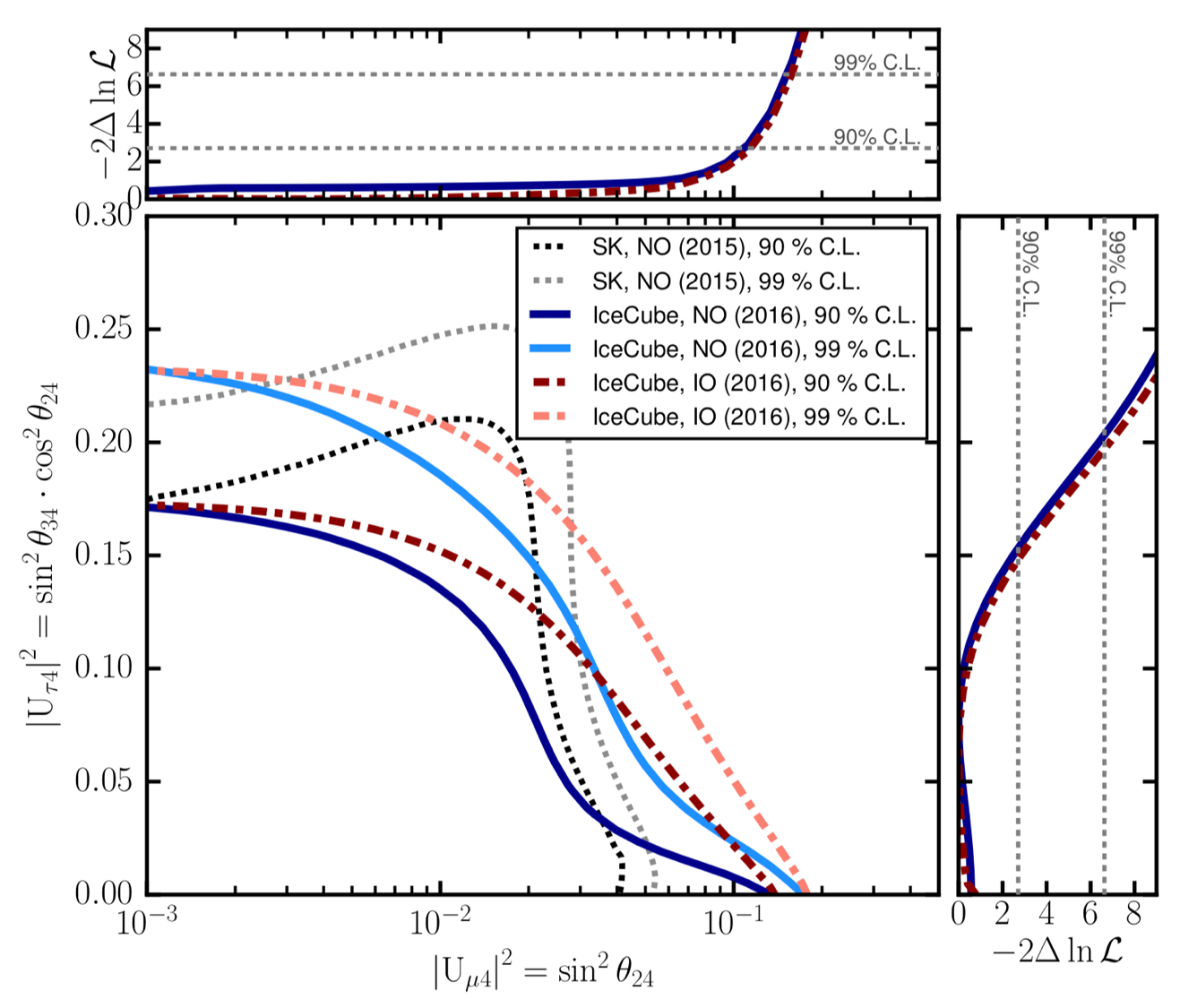}
 
\caption{Results from the IceCube low-energy sterile neutrino search assuming each mass ordering, compared to similar results from SuperKamiokande. \label{fig:LEResult}}
\end{center}
\end{figure}

No evidence of atmospheric muon neutrino disappearace was observed, leading to a limit expressed in terms of the mixing matrix elements $|U_{\mu 4}|^2=\sin^2 \theta_{24}$ and $|U_{\tau 4}|^2=\sin^2 \theta_{34}\cos^2 \theta_{34}$, shown in Fig.~\ref{fig:LEResult}

\section{Future Plans}
\label{sec-3}
The IceCube collaboration is preparing an extended, 7 year high-energy sterile neutrino search. The event selection has been enhanced, with increased efficiency, especially at low-energy, while retaining an effectively background-free selection.  The number of expected event is approximately 280,000, which is 13 times as many as in the published one-year analysis.  With this enhancement of  statistical precision comes a need for corresponding control of systematic uncertainties, which have undergone significant improvements for this analysis.

Bulk ice uncertainties arising from the depth-dependent dust distribution within the IceCube detector have been studied using a multidimensional procedure producing covariance matrices in analysis space. These matrices encode the ice model variability allowed within constraints from LED calibration data, complete with all depth-dependent correlations, extending beyond the effective uncertainty on global absorption and scattering coefficients used in previous analyses.  A continuous parameterizartion of refrozen hole ice scattering has been incorporated, deriving from advances in understanding its properties from lower energy analyses.  An advanced treatment of the atmospheric flux parameters has been implemented using the ``Barr scheme'' \cite{Barr:2006it}, applying 6 effective parameters that capture the uncertainty in hadronic modelling of the air shower, constrained by collider data. This replaces the effective $\pi / K$ and $\nu /\bar \nu$ parameters with a more physically motivated and complete uncertainty parameterization. Finally, the next generation of the sterile neutrino search will treat the effects of non-zero $\theta_{34}$ explicitly, to provide confidence intervals in $\theta_{24}$ and $\Delta m^2$ at several fixed $\theta_{34}$ points, rather than simply providing the most conservative limit at $\theta_{34}=0$.  

The next-generation high-energy sterile neutrino search at IceCube is in development and will have unprecedented sensitivity to muon neutrino disappearance using high-energy atmospheric neutrinos.  Observation of such a disappearance signature would represent a major discovery.   If no disappearance is observed, however, further severe limitations will be placed on sterile neutrino models to explain the short baseline neutrino anomalies. 

\bibliography{biblio}

\end{document}